This work was carried out in whole or in part within the framework of the NOMATEN Centre of Excellence, supported from the European Union Horizon 2020 research and innovation program (Grant Agreement No. 857470) and from the European Regional Development Fund via the Foundation for Polish Science International Research Agenda PLUS program (Grant No. MAB PLUS/2018/8), and the Ministry of Science and Higher Education's initiative "Support for the Activities of Centers of Excellence Established in Poland under the Horizon 2020 Program" (agreement no. MEiN/2023/DIR/3795).

The version of record of this article, first published in International Journal of Refractory Metals and Hard Materials, Volume 139, September 2026, 107772, is available online at Publisher's website: https://doi.org/10.1016/j.ijrmhm.2026.107772

# Nanoindentation induced plasticity in equiatomic MoTaW alloys by experimentally guided machine learning molecular dynamics simulations

F. J. Dominguez-Gutierrez[1], T. Stasiak[1], G. Markovic[2], A. Kosinska[1], K. Mulewska[1]

[1]*National Centre for Nuclear Research, NOMATEN CoE, ul. Andrzeja Soltana 7, 05-400 Świerk, Poland*
[2]*Institute for Technology of Nuclear and Other Mineral Raw Materials, 11000 Belgrade, Serbia*

## Abstract

Refractory complex concentrated alloys (RCCA) exhibit exceptional strength and thermal stability, yet their plastic deformation mechanisms under complex contact loading remain insufficiently understood. Here, the nanoindentation response of an equiatomic MoTaW alloy is investigated through a combined experimental and atomistically resolved modeling approach. Spherical nanoindentation experiments are coupled with large-scale molecular dynamics simulations employing a tabulated low-dimensional Gaussian Approximation Potential (tabGAP), enabling near-density-functional-theory accuracy at experimentally relevant length scales. A physics-based similarity criterion, implemented via principal component analysis of load–displacement curves, is used to identify mechanically representative experimental responses for quantitative comparison with simulations. Indentation stress–strain curves are constructed yielding a good agreement in the elastic regime between experiment and simulation, with reduced Young's moduli of approximately 270 GPa. Generalized stacking fault energy calculations reveal elevated unstable stacking- and twinning-fault energies in MoTaW relative to pure refractory elements, indicating suppressed localized shear and a preference for dislocation-mediated plasticity. Atomistic analyses demonstrate a strong crystallographic dependence of plastic deformation, with symmetric $\{110\}\langle 111\rangle$ slip activation and four-fold rosette pile-ups for the [001] orientation, and anisotropic slip, strain localization, and enhanced junction formation for [011]. Local entropy and polyhedral template matching analyses further elucidate dislocation network evolution and deformation-induced local structural transformations. Overall, this study establishes a direct mechanistic link between fault energetics, orientation-dependent dislocation activity, and experimentally observed nanoindentation behavior in RCCA.

## 1. Introduction

Refractory complex concentrated alloys (RCCAs) represent a class of alloys composed of three or more principal refractory elements, typically including Mo, Ta, W, Nb, and Hf, and are designed to achieve high strength and phase stability at elevated temperatures [1-3]. Due to the predominance of body-centered cubic (BCC) crystal structures, many RCCAs exhibit superior high-temperature strength compared to conventional refractory alloys and dilute solid solutions [1,4, 5]. Existing experimental and computational studies have largely focused on Nb-containing RCCA systems, such as MoNbTaW, NbTiZrV, NbTiZrCr, and HfNbTaTiZr alloys, which have been extensively investigated with respect to their high-temperature mechanical behavior, oxidation resistance, and defect- and impurity-related effects in chemically complex BCC matrices [5–7]. In contrast, Nb-free refractory compositions have received comparatively limited attention in bulk form. Among these, the ternary Mo–Ta–W system is a refractory alloy composed exclusively of Group V and VI elements. Previous investigations of Mo–Ta–W alloys, particularly in thin-film form, have demonstrated high hardness, elastic modulus, and structural stability, highlighting the intrinsic mechanical robustness of this system and motivating further studies of its deformation behavior under complex loading conditions [8].

Atomistic simulations are a powerful tool for understanding the relationships between composition, structure, and mechanical behavior in complex alloys; however, their predictive capability critically depends on the quality of the employed interatomic potentials [9-14]. For chemically complex systems such as high-entropy and refractory complex concentrated alloys, the development of reliable empirical interatomic potentials is particularly challenging, as predefined functional forms often fail to capture the diversity of local atomic environments [10, 15-17]. Machine-learning interatomic potentials (MLIPs) provide an alternative framework in which the potential-energy surface is not assumed *a priori*, but is instead learned directly from quantum-mechanical reference data, typically obtained from density functional theory calculations [9,16]. As a result, MLIPs enable molecular dynamics (MD) simulations with near-DFT accuracy while retaining computational efficiency sufficient to access deformation mechanisms, defect evolution, and mechanical responses at time and length scales inaccessible to *ab initio* methods [15-18]. Importantly, recent studies demonstrate that MLIPs can reliably capture defect energetics and deformation mechanisms in chemically complex alloys, including stacking fault formation and dislocation activity, which are essential for interpreting mechanical behavior beyond the elastic regime [17].

Among the various MLIP formalisms, Gaussian Approximation Potentials (GAPs) provide a flexible framework in which the total energy is expressed as a sum of local atomic contributions described by physically motivated descriptors. However, for chemically complex refractory alloys, the use of high-dimensional descriptors can become computationally prohibitive, particularly for large-scale molecular dynamics simulations. To address this limitation, low-dimensional GAP formulations combined with tabulation strategies (tabGAP) have been introduced, enabling efficient evaluation of energies and forces via spline interpolation while retaining near-DFT accuracy for bulk properties and defect energetics [18,19]. This approach has been successfully applied to refractory bcc high-entropy and complex concentrated alloys, including Mo–Nb–Ta–V–W-based systems, demonstrating its suitability for capturing defect behavior, phase stability, and deformation-relevant energetics in multicomponent refractory alloys [18-22]. In this context, a detailed understanding of the deformation mechanisms governing Nb-free refractory alloys under localized contact loading is still lacking. MD simulations are widely employed to investigate the atomistic mechanisms governing deformation and defect evolution in crystalline materials under mechanical loading. By explicitly resolving atomic trajectories and interatomic interactions, MD provides direct access to incipient plasticity, dislocation nucleation, and local stress–strain responses at nanometer scales, complementing experimental nanoindentation measurements where these processes are challenging to isolate. This multiscale capability has been successfully applied across a range of material systems to elucidate fundamental deformation pathways and inform constitutive models [23]. In the present study, MD serves as a deterministic reference for interpreting orientation-dependent plasticity and pile-up morphology observed experimentally. In particular, the relationship between atomistic fault energetics, crystallographic orientation, and the resulting plastic response during nanoindentation remains insufficiently explored in multicomponent refractory systems. Addressing this gap requires a combined experimental–computational approach capable of resolving dislocation-mediated plasticity across multiple length scales.

In this work, we investigate the mechanical response of an equiatomic MoTaW refractory alloy by combining spherical nanoindentation experiments with large-scale molecular dynamics simulations based on a tabulated low-dimensional Gaussian Approximation Potential (tabGAP). Generalized stacking fault energies are first evaluated to characterize the energetic barriers governing slip and twinning in the alloy. These atomistic insights are then directly correlated with orientation-dependent plasticity observed during nanoindentation, enabling a mechanistic interpretation of dislocation nucleation, propagation, and strain localization under contact loading. By explicitly linking GSFE-derived fault energetics with experimentally observed indentation behavior, the present study provides an atomistically informed framework for understanding contact-induced plasticity in refractory complex concentrated alloys.

## 2. Methods

.

### 2.1 Experimental methods

Equiatomic MoTaW refractory complex concentrated alloys samples were synthesized using an arc-melting technique in an Arc Melter AM 200 system (Edmund Bühler, Germany). The starting materials consisted of commercially available elemental pellets of Mo, Ta, and W (Kurt J. Lesker, UK and Goodfellow, UK), each with a purity exceeding 99.90 wt.%. Prior to melting, the chamber was evacuated to a base pressure of approximately $10^{-5}$ mbar using a turbomolecular pump to minimize oxidation. Melting was subsequently carried out under a high-purity argon atmosphere at a partial pressure of 0.5 mbar [5]. To further reduce residual oxygen in the chamber, a titanium getter was melted prior to alloy synthesis. Button-shaped alloy samples with masses ranging from 17 to 20 g were produced. To ensure chemical homogeneity, each sample was remelted ten times, with the button flipped between successive melts. The produced samples were submitted to homogenization annealing in a Czylok vacuum furnace (Czylok, Poland) at 1200 °C for 24 hours. The heating was performed at a rate of 10 °C per minute, while cooling was carried out slowly in the furnace. The resulting MoTaW alloy exhibited a uniform elemental distribution of approximately 33.3 at.% for each constituent element. The alloy consists of a single bcc structure and large grains of several hundred micrometers [6]. The microstructural characterization of the homogenized sample was performed using a ThermoFisher Helios 5 UX scanning electron microscope equipped with an EDAX Velocity Pro EBSD detector for high-resolution crystallographic orientation mapping. Electron Backscatter Diffraction (EBSD) data were acquired with a step size of 0.25 µm. The collected orientation maps were analyzed using EDAX OIM Analysis 8 software. Data points with a confidence index (CI) lower than 0.1 were excluded from the analysis to ensure reliability of the crystallographic information. Nanoindentation experiments were performed at room temperature using a NanoTest Vantage system (Micro Materials Ltd.) equipped with a spherical diamond indenter tip of 5 µm radius. Indentations were conducted in single-force mode with a maximum applied load of 15 mN. Loading was applied at a constant rate of 1 mN/s, followed by a 5 s hold segment at peak load, and unloading at a rate of 3 mN/s. The spacing between indents was 50 µm to avoid interaction between plastic zones. The maximum penetration depth reached approximately 150 nm. Prior to testing, the diamond area function (DAF) for the spherical indenter was calibrated using fused silica with known mechanical properties. Contact depth, hardness, and reduced modulus were extracted using the standard Oliver–Pharr analysis framework.

## 2.2 Computational modeling

In this work, interatomic interactions are described by using recently developed Machine Learning Interatomic Potential that is based on the tabulated implementation of a low-dimensional Gaussian approximation potential (tabGAP) [21,25]. In the GAP formalism, the total energy of the system is expressed as a sum of local atomic energy contributions that depend on the surrounding atomic environments and are fitted to density functional theory (DFT) reference data [26,27]. In contrast to high-dimensional descriptor approaches such as SOAP, the low-dimensional GAP framework employed here represents atomic environments using a limited set of physically motivated descriptors, including pairwise distance terms, angular three-body correlations, and a scalar many-body density contribution analogous to the embedded-atom method [26]. This reduced descriptor dimensionality preserves the essential physics of metallic bonding namely, coordination dependence, angular correlations, and local density effects while enabling a compact representation of the potential energy surface. The tabGAP formulation exploits this low-dimensional representation by pre-tabulating the descriptor-dependent energy contributions on regular grids and evaluating them during molecular dynamics simulations via spline interpolation [22,25]. As a result, the computational cost of force and energy evaluations is dramatically reduced compared to conventional GAP implementations, while maintaining near-DFT accuracy across crystalline, defected, and chemically disordered configurations [22]. The tabulated form also ensures numerical stability and smooth force fields, which are critical for long-time molecular dynamics simulations and deformation studies. Owing to this balance between physical fidelity, numerical robustness, and computational efficiency, TabGAP is particularly well suited for modeling refractory bcc metals and multicomponent alloys, where complex local chemical environments and large simulation cells are required to capture mechanisms such as generalized stacking fault energetics, dislocation nucleation, and nanoindentation-induced plasticity [11,13, 28, 29].

Generalized stacking fault energy (GSFE) curves were computed by constructing simulation cells in which the crystallographic orientation is chosen such that the target glide plane coincides with the *x*–*y* plane and the shear direction lies along ⟨111⟩. For the {110}⟨111⟩ shear path, the simulation cell was defined using lattice vectors oriented along ⟨1 −1 1⟩ in the *x* direction, ⟨1 −1 −2⟩ in the *y* direction, and ⟨110⟩ in the *z* direction, such that the {110} plane normal aligns with the *z* axis [27-29]. For the {112}⟨111⟩ case, the lattice vectors were chosen as ⟨1 1 −1⟩ along *x*, ⟨−1 1 0⟩ along *y*, and ⟨112⟩ along *z*, placing the {112} plane normal along the *z* direction [29]. In both configurations, the ⟨111⟩ direction lies within the fault plane and defines the shear direction. Each oriented unit cell was replicated to form a 12 × 10 × 5 supercell, resulting in a system containing 3600 atoms. A vacuum region of approximately 2.0 nm was introduced along the *y* direction to eliminate spurious interactions between periodic images across the fault plane, consistent with established GSFE methodologies. GSFE curves were generated by incrementally displacing the atoms in the upper half of the crystal relative to the lower half along the ⟨111⟩ shear direction [13, 15, 30, 31]. The displacement was applied in uniform increments corresponding to 0.1 of the Burgers vector magnitude, spanning a full lattice periodicity. Following each displacement step, the atoms in the top and bottom layers were held fixed, while the remaining atoms were allowed to relax only along the direction normal to the fault plane [13, 31]. Structural relaxation was performed via conjugate-gradient energy minimization and was deemed converged when either (i) the relative change in total energy between successive iterations fell below $10^{-12}$ or (ii) the global force norm acting on all atoms was less than $10^{-12}$ eV Å$^{-1}$. Then, the stacking fault energy can be calculated as:

$$\gamma_{GSFE} = \frac{E_S - E_0}{A_{SF}} \quad (1)$$

where $E_s$ represents the energy of the sample at a given displacement, and $E_0$ denotes the energy for the perfect sample, $A_{SF}$ stands for the stacking fault area. For the pure elemental systems, W, Mo, and Ta were modeled using their respective equilibrium lattice constants, taken as 3.165 Å for W, 3.160 Å for Mo, and 3.301 Å for Ta, consistent with experimental and first-principles data [13, 20, 32]. The Mo–Ta–W ternary alloy was generated starting from an initial bcc Mo supercell with a lattice constant of 3.20 Å, chosen to approximate the average lattice spacing expected from Vegard's law. In this construction, 66% of the Mo atoms were randomly substituted by W and Ta atoms in equiatomic proportions, yielding a chemically disordered MoTaW configuration while preserving the underlying bcc lattice topology. Following chemical substitution, all alloy configurations were fully relaxed using the FIRE (Fast Inertial Relaxation Engine) 2.0 algorithm [33]. to obtain the nearest local minimum of the potential energy surface. Geometry optimization was considered converged when both of the following criteria were satisfied: (i) the relative change in total energy between successive iterations fell below $10^{-5}$, and (ii) the global norm of the force vector acting on all atoms was less than $10^{-8}$ eV Å$^{-1}$ [12, 13]. This relaxation procedure ensures that the effects of local lattice distortions induced by chemical disorder are fully accounted for prior to the computation of generalized stacking fault energies [12].

For the modeling of nanoindentation in the MoTaW alloy, large-scale simulation cells were initially constructed using pure bcc Mo as a reference lattice, owing to its structural compatibility with both W and Ta and its intermediate lattice parameter [13, 32]. The dimensions of the simulation cells were selected to accommodate the expected nucleation and propagation of dislocation loops along dominant bcc slip families, while minimizing finite-size effects and interactions with the simulation boundaries. Two crystallographic surface orientations were investigated. In all cases, the simulation cell was defined such that the z-axis coincides with the indentation loading direction and therefore represents the surface normal, while the x- and y-axes define the in-plane directions of the free surface. For the [001] orientation, the loading direction (z-axis) is parallel to [001], with the in-plane axes aligned along [100] and [010], corresponding to a (001) surface. For the [011] orientation, the z-axis is aligned with [011], and the in-plane directions are defined by [100] and [0 1 −1], resulting in a (011) surface. This explicit definition of the crystallographic basis clarifies the loading geometry and the projection of the active {110}⟨111⟩ slip systems onto the free surface. For the [001] orientation, the oriented Mo unit cell was replicated (156 × 158 × 160) times along the Cartesian directions, resulting in a system containing 7887360 atoms, while for the [011] orientation, the unit cell was replicated (160 × 120 × 110) times, yielding 8448000 atoms. The anisotropic replication factors were chosen to ensure comparable lateral dimensions and sufficient depth across all orientations, thereby allowing for the unrestricted development of indentation-induced plastic zones. Following construction of the pure Mo simulation cells, the MoTaW alloy configurations were generated by random substitution of Mo atoms with W and Ta atoms, as described for the GSFE calculations, yielding an equiatomic chemical composition while preserving the underlying bcc lattice topology. Subsequent full structural relaxation was performed prior to nanoindentation to account for local lattice distortions induced by chemical disorder. This unified construction strategy ensures consistency between the GSFE and nanoindentation simulations, enabling a direct correlation between fault energetics and indentation-induced deformation mechanisms. The x and y axes have periodic boundary conditions to simulate an infinite surface, and each sample was divided along the z-axis. The bottom 2% of the thickness was fixed, and the next 8% served as a thermostatic layer to dissipate indentation heat. The remaining atoms formed the dynamic region for atoms–indenter interaction allowing atoms to move freely, as shown in Figure 1. Additionally, a 5 nm vacuum region is added atop the sample as an open boundary [12, 14]. The samples are initially equilibrated at T = 300 K for 2 ns using an isobaric-isothermal ensemble. This equilibration is achieved by integrating the Nose–Hoover equations with damping parameters, specifically, $\tau_T = 2$ fs for the thermostat and $\tau_P = 5$ ps for the barostat, while maintaining an external pressure of 0 GPa. This process continues until the system attains homogeneous temperature and pressure profiles [23].

Indentation was carried out using a rigid spherical indenter with force $F(t) = K\,(r(t) - R)^2$, where K = 200 eV/Å³ is the force constant and R = 15 nm is the indenter radius. The indenter center followed $r(t) = (x_0, y_0, z_0 \pm vt)$ with an initial offset $z_0 = 0.5$ nm and indentation velocity of v = 20 m/s. Periodic boundary conditions were applied along x and y, and the indentation depth was limited to 4 nm to reduce boundary effects [12,14, 32]. 10 MD simulations are performed considering different locations on the surface and each simulation lasts 225 ps with a timestep of 1 fs. The indentation velocity used in the MD simulations (20 m $s^{-1}$) is several orders of magnitude higher than in experiments, which is a well-known and unavoidable limitation of atomistic simulations due to accessible time scales. Importantly, this velocity remains well below the longitudinal speed of sound in bcc refractory metals (≈4–6 km $s^{-1}$ for Mo, Ta, and W), ensuring that deformation proceeds in a quasi-static regime without shock-wave effects. While elevated loading rates can increase the apparent stress required for dislocation nucleation due to reduced thermal activation, previous studies have shown that relative trends in slip activation, pile-up morphology, and orientation dependence are largely insensitive to indentation rate [13]. In this work, the MD simulations are therefore used to capture the dominant crystallographic deformation mechanisms rather than absolute critical stresses. The load–displacement curve was obtained by plotting the indenter force against penetration depth reporting the average of 10 MD simulations [11]. All MD simulations are performed using Large-scale Atomic/Molecular Massively Parallel Simulator (LAMMPS) software, which allows us to study the behavior of materials under a wide range of conditions [34]. Dislocation structures were analyzed using OVITO [36] and identified by the Dislocation Extraction Algorithm (DXA) [36]. In addition, structural identification was performed using the Polyhedral Template Matching (PTM) algorithm implemented in OVITO [35]. The analysis was restricted to BCC, FCC, and HCP template structures to avoid overclassification of distorted environments, and a global RMSD cutoff of 0.10 was applied. Atoms with RMSD values exceeding this threshold were classified as non-crystalline (defect or disordered regions). This cutoff was selected following standard practice to provide a robust distinction between thermally perturbed lattice atoms and atoms associated with defects, while minimizing false identification [12].

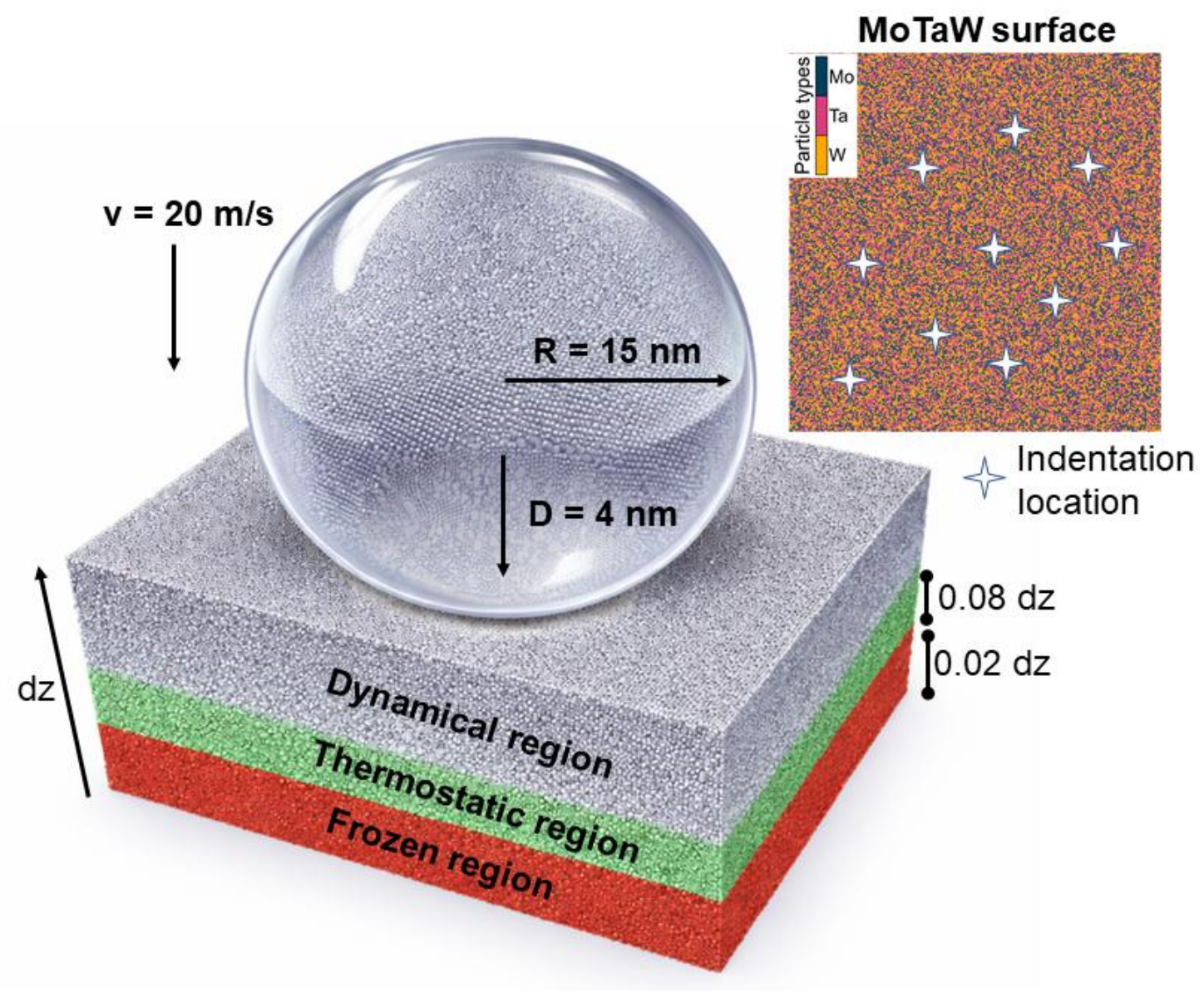

Fig 1. Schematics of the sample preparation for nanoindentation simulations.

To enable a rigorous comparison between experimental nanoindentation data and deterministic MD simulations, a physics-guided machine learning (ML) framework was developed to extract mechanically representative load–displacement (LD) curves from the complete dataset (200 indents). All experimental LD curves P(h) were resampled onto a common penetration-depth grid using piecewise linear interpolation. Extrapolation beyond the measured depth range was not permitted; values outside the original data domain were assigned as undefined. To ensure sufficient data coverage and exclude truncated or incomplete responses, only curves with at least 70% finite interpolated points on the common grid were retained for further analysis. Normalization was performed exclusively on the valid (finite) portion of each retained curve. A baseline correction was first applied by subtracting the minimum load value to eliminate small offsets associated with contact detection or instrumental noise. The corrected curve was then scaled by its maximum load so that each LD response was normalized to a unit peak load. Curves exhibiting non-physical behavior, such as zero or negative load range after baseline correction, were discarded. Undefined regions were preserved in the vector representation to prevent artificial data generation during interpolation. This procedure ensures that all LD curves are represented on a consistent depth grid while retaining physically meaningful shape descriptors, such as elastic slope, elastic–plastic transition behavior, and post-yield curvature, required for subsequent principal component analysis and robust similarity filtering. Each curve was then represented as a vector in a high-dimensional space defined by the discretized depth values. Principal Component Analysis (PCA) was applied to reduce dimensionality where the first three principal components captured more than 95% of the total variance and were retained to construct a reduced mechanical feature space. In this representation, the primary components reflect variations in elastic slope, elastic–plastic transition, and plastic curvature evolution. A robust similarity metric was defined in the reduced PCA space. The median response vector was computed, and for each curve a robust Z-score was evaluated using the median absolute deviation (MAD):$Z_i = \frac{|| X_i - X_{median} ||}{MAD}$. A strict threshold ($|Z| \leq 0.07$) was imposed to retain only curves exhibiting identical elastic response and consistent plastic evolution within the intrinsic experimental repeatability envelope. Curves exceeding this threshold were identified as mechanically distinct and excluded. This approach constitutes a physics-guided dimensionality reduction and robust similarity analysis framework tailored for spherical nanoindentation datasets, enabling objective identification of the intrinsic mechanical response for quantitative comparison with atomistic simulations.

## 3. Results

Figure 2a) presents the GSFE curves for the $\{110\}\langle 111\rangle$ slip system in W, Mo, Ta, and the equiatomic MoTaW alloy. For all materials, the GSFE exhibits a single pronounced maximum at approximately half of the lattice translation, corresponding to the unstable stacking fault energy γusf, which represents the energetic barrier for homogeneous shear and serves as a key descriptor for dislocation nucleation [27]. Among the investigated systems, W displays the highest $\gamma_{usf}$, with a value of 1.676 J m$^{-2}$, consistent with its well-known high lattice resistance to dislocation motion and large Peierls barrier [11, 26, 37]. Mo shows a slightly lower $\gamma_{usf}$ of 1.475 J m$^{-2}$, reflecting its comparatively lower resistance to shear while retaining a similar qualitative GSFE shape [13, 31]. In contrast, Ta exhibits a substantially reduced $\gamma_{usf}$ of 0.737 J m$^{-2}$, in agreement with its softer shear response and lower intrinsic lattice friction among bcc refractory metals [31, 36]. The equiatomic MoTaW alloy displays an intermediate $\gamma_{usf}$ of 1.361 J m$^{-2}$, lying between those of W and Mo and significantly above that of Ta. Notably, the $\gamma_{usf}$ of MoTaW exceeds the simple arithmetic average of its constituent elements, indicating a non-linear energetic response associated with chemical disorder and local lattice distortion. This increase in the unstable fault energy suggests that, despite the presence of the low-barrier Ta component, shear in MoTaW is governed by the stronger local bonding environments characteristic of Mo- and W-rich atomic configurations [13,28,31]. From a mechanical perspective, the elevated $\gamma_{usf}$ in MoTaW implies a higher barrier for dislocation nucleation compared to pure Ta, which is expected to contribute to increased hardness and delayed onset of plasticity under nanoindentation [39]. At the same time, the reduced $\gamma_{usf}$ relative to pure W indicates that chemical disorder partially mitigates the extreme lattice resistance of tungsten, potentially facilitating more distributed plastic deformation once dislocation activity is initiated. These trends are fully consistent with the experimentally observed balance between high strength and non-catastrophic plastic flow in refractory multicomponent alloys.

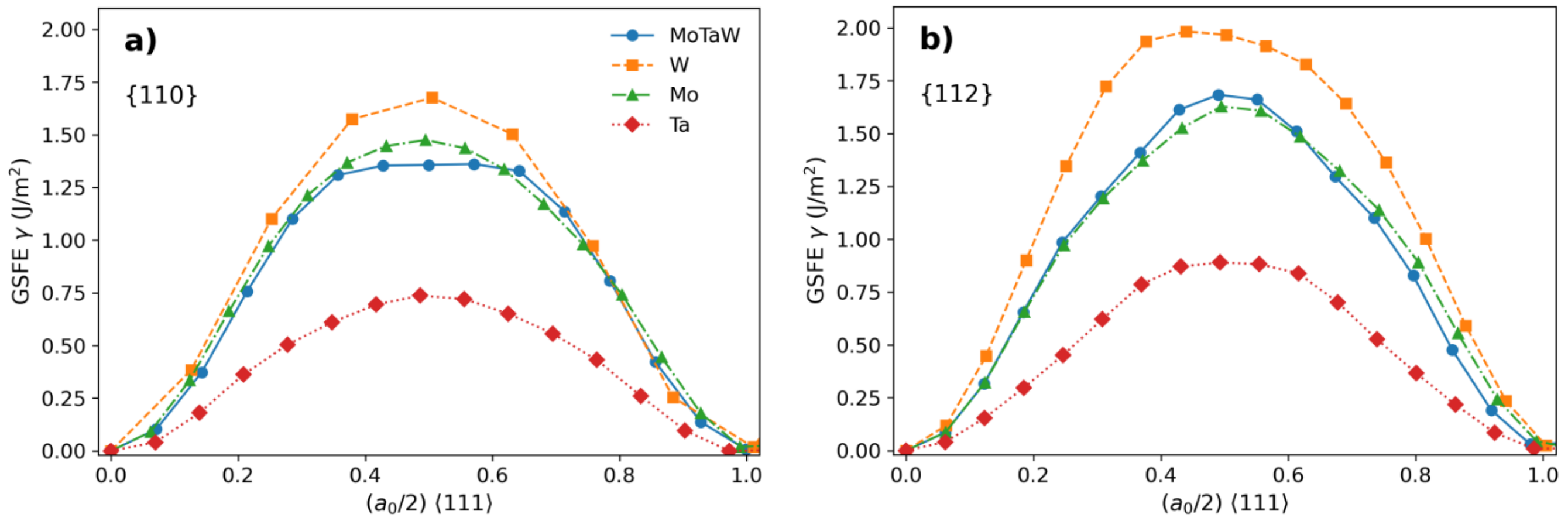

Fig 2. Generalized stacking fault energy (GSFE) curves for bcc W, Mo, Ta, and equiatomic MoTaW computed using the TabGAP machine-learning interatomic potential: (a) {110}⟨111⟩ slip system and (b) {112}⟨111⟩ shear path associated with twinning. Displacements are normalized by a0/2 along the ⟨111⟩ direction, and energies are reported per unit fault area.

Figure 2b) shows the GSFE curves for the {112}⟨111⟩ shear path in W, Mo, Ta, and the equiatomic MoTaW alloy. In bcc crystals, this slip system is closely associated with twinning-related shear, and the maximum of the GSFE curve is commonly interpreted as the unstable twinning-fault energy, $\gamma_{utf}$, representing the energetic barrier for the nucleation of a twin embryo. Among the investigated materials, W exhibits the highest $\gamma_{utf}$, with a value of 1.983 J m$^{-2}$, reflecting its strong directional bonding and pronounced resistance to twinning-related shear [28,37]. Mo shows a slightly lower $\gamma_{utf}$, of 1.628 J m$^{-2}$, indicating a reduced but still substantial barrier for twin nucleation [13,31]. In contrast, Ta displays a markedly lower $\gamma_{utf}$ of 0.889 J m$^{-2}$, consistent with its well-known tendency toward easier plastic deformation and enhanced twinning activity under complex stress states [32,36]. The equiatomic MoTaW alloy exhibits an intermediate $\gamma_{utf}$ of 1.683 J m$^{-2}$, which lies between those of W and Mo and is significantly higher than that of Ta. Notably, the γutf of MoTaW exceeds the simple rule-of-mixtures expectation, highlighting a non-linear energetic response induced by chemical disorder and local lattice distortions. Despite the presence of the low-barrier Ta component, the twinning-fault barrier in MoTaW remains governed by the stronger local bonding environments associated with Mo- and W-rich atomic neighborhoods. A comparison between Figures (a) and (b) further reveals that, for all materials, the {112}⟨111⟩ shear path is associated with a higher energetic barrier than the corresponding {110}⟨111⟩ slip system, underscoring the intrinsic preference for slip over twinning in bcc refractory metals. However, in MoTaW, the difference between $\gamma_{usf}$ ({110}) and $\gamma_{utf}$ ({112}) is reduced compared to pure W, suggesting a moderation of slip–twinning anisotropy due to chemical complexity. This reduction in anisotropy is expected to promote the activation of competing deformation mechanisms under multiaxial loading, such as nanoindentation. From a mechanical standpoint, the elevated $\gamma_{utf}$ in MoTaW implies a suppressed propensity for deformation twinning, favoring dislocation-mediated plasticity at the onset of yielding. At the same time, the intermediate magnitude of $\gamma_{utf}$ relative to W indicates that chemical disorder partially alleviates the extreme lattice resistance characteristic of pure tungsten, potentially contributing to the experimentally observed balance between high strength and distributed plastic flow in refractory multicomponent alloys. To validate the reliability of the newly developed TabGAP MLIP for refractory alloys, all extracted GSFE quantities including the unstable stacking-fault energy ($\gamma_{usf}$) for the {110}⟨111⟩ system and the unstable twinning-fault energy ($\gamma_{utf}$) for the {112}⟨111⟩ system were systematically tabulated and compared with values reported in the literature in Tab1. These include results obtained using DFT, EAM and SNAP for bcc W, Mo, and Ta; while rule of mixtures was used for the MoTaW alloy. These results demonstrate that the TabGAP potential accurately captures the energetics governing slip and twinning in refractory bcc systems, thereby providing a robust foundation for the atomistic interpretation of nanoindentation experiments and complex deformation mechanisms in MoTaW alloys.

Tab 1. unstable stacking fault energy on the {110} and unstable twinning fault energy on the {112} plane in J/m$^2$

| Parameter | MoTaW | W | Mo | Ta | IP |
|---|---|---|---|---|---|
| $\gamma_{usf}$ | 1.361 | 1.676 | 1.475 | 0.737 | TabGAP [12] |
| $\gamma_{usf}$ | 1.413 | 1.910 | 1.499 | 0.831 | SNAP [38] |
| $\gamma_{usf}$ | 1.313 | 1.772 | 1.443 | 0.725 | DFT [39] |
| $\gamma_{usf}$ | 1.316 | 1.740 | 1.458 | 0.751 | EAM [14] |
| $\gamma_{utf}$ | 1.683 | 1.983 | 1.628 | 0.889 | TabGAP [12] |
| $\gamma_{utf}$ | 1.578 | 2.121 | 1.636 | 0.978 | SNAP [38] |
| $\gamma_{utf}$ | 1.383 | 1.846 | 1.465 | 0.838 | DFT [39] |
| $\gamma_{utf}$ | 1.523 | 2.011 | 1.689 | 0.868 | EAM [14] |

### 3.1. Load–displacement response and incipient plasticity

To enable a meaningful comparison between experimental nanoindentation data and deterministic MD simulations, the experimental load–displacement (LD) curves were first filtered to identify a mechanically representative subset. All recorded LD curves were projected onto a low-dimensional space using principal component analysis (PCA), which captures the dominant variance associated with elastic stiffness, the elastic–plastic transition, and subsequent plastic deformation [42]. In this reduced space, curves corresponding to the same underlying deformation mechanism form a compact cluster, while curves affected by experimental artifacts, surface imperfections, or premature failure appear as statistical outliers [43]. A robust similarity criterion was applied by measuring the distance of each curve from the median response in the PCA space [43]. Only curves lying within a narrow threshold, Z-score range less than 0.07, were retained, as shown in Fig 3a). The threshold was selected to impose a narrow similarity window in the reduced feature space, thereby retaining only curves whose elastic slope, elastic–plastic transition, and post-yield curvature remain within the intrinsic experimental repeatability envelope. Importantly, this selection is physics-based rather than statistical, as it isolates curves that share the same elastic response, onset of plasticity, and overall deformation pathway, rather than merely minimizing numerical variance [43,44]. As depicted in Fig 3b), the distribution of robust z-scores, shown as a histogram, reveals a sharply peaked central population corresponding to mechanically consistent load–displacement responses, surrounded by a small number of outliers with significantly larger deviations. Specifically, this range corresponds to deviations that are comparable to the intrinsic experimental scatter arising from instrument noise, surface roughness, and minor variations in tip–sample contact conditions, while excluding curves affected by clear experimental anomalies such as pop-in instability, surface defects, or partial tip–surface misalignment [44]. The choice of this threshold was guided by the observed spread of the robust z-score histogram, which exhibits a pronounced central cluster associated with highly reproducible elastic–plastic responses and long tails corresponding to outlier curves. Selecting $|Z| \leq 0.07$ resulted in the retention of 10 out of 200 experimental load–displacement curves, corresponding to 5% of the dataset, this ensures retention of curves whose normalized shapes remain within the experimental repeatability envelope of the nanoindentation system, as defined by the calibration uncertainty and load–displacement resolution, while rejecting responses that deviate beyond physically meaningful variability. The remaining 95% were excluded as mechanically distinct from the median response in the reduced feature space. It is important to emphasize that this high exclusion rate does not indicate poor data quality; rather, it reflects the intrinsic variability of localized nanoindentation measurements in chemically complex, large-grained materials, where small differences in surface condition, local orientation deviation, and microstructural heterogeneity can produce measurable variations in load–displacement response. The applied threshold therefore isolates a compact subset of curves that share an identical deformation pathway, enabling a rigorous and deterministic comparison with atomistic simulations. This physics-informed filtering strategy avoids arbitrary data selection and yields a compact yet statistically representative dataset that faithfully captures the intrinsic mechanical response of the MoTaW alloy, enabling a consistent and meaningful comparison with atomistic nanoindentation simulations.

The indentation stress–strain response was computed within the classical spherical contact framework originally established by Hertz [45] for elastic contact mechanics and later extended to instrumented indentation analysis in comprehensive reviews by Lawn [46] and Fischer-Cripps [48]. Building upon this foundation, we adopt the formulation proposed by Pathak and Kalidindi [48,49], which establishes a physically consistent mapping between spherical nanoindentation measurements and an equivalent uniaxial stress–strain description of the material response. In this approach, the indentation stress is defined as the mean contact pressure beneath the indenter, σ=P/A, while the indentation strain is expressed as a normalized measure of deformation, ε=a/R, where P is the applied load, A is the projected contact area, $a^2 = 2hcR - hc^2$ is the contact radius, and R is the indenter radius. For the experimental data, the contact radius and area were obtained from the contact depth hc, which was calculated using the Oliver–Pharr relation hc=ht−3P/4S, where ht is the maximum penetration depth and S is the unloading stiffness measured at peak load [50, 51]. This formulation accounts for elastic recovery during unloading and enables an accurate estimation of the contact geometry. In contrast, for MD simulations, unloading information is generally unavailable; therefore, the contact area was computed directly from the exact spherical geometry of the indenter using the instantaneous penetration depth [12,14]. This purely geometric definition is consistent with the assumptions of the Kalidindi framework and avoids ambiguities associated with stiffness extraction in atomistic systems [14]. Despite these methodological differences, both experimental and MD stress–strain curves are defined on the same physical basis, allowing for direct comparison of elastic response, yield onset, and subsequent plastic deformation mechanisms across length scales.

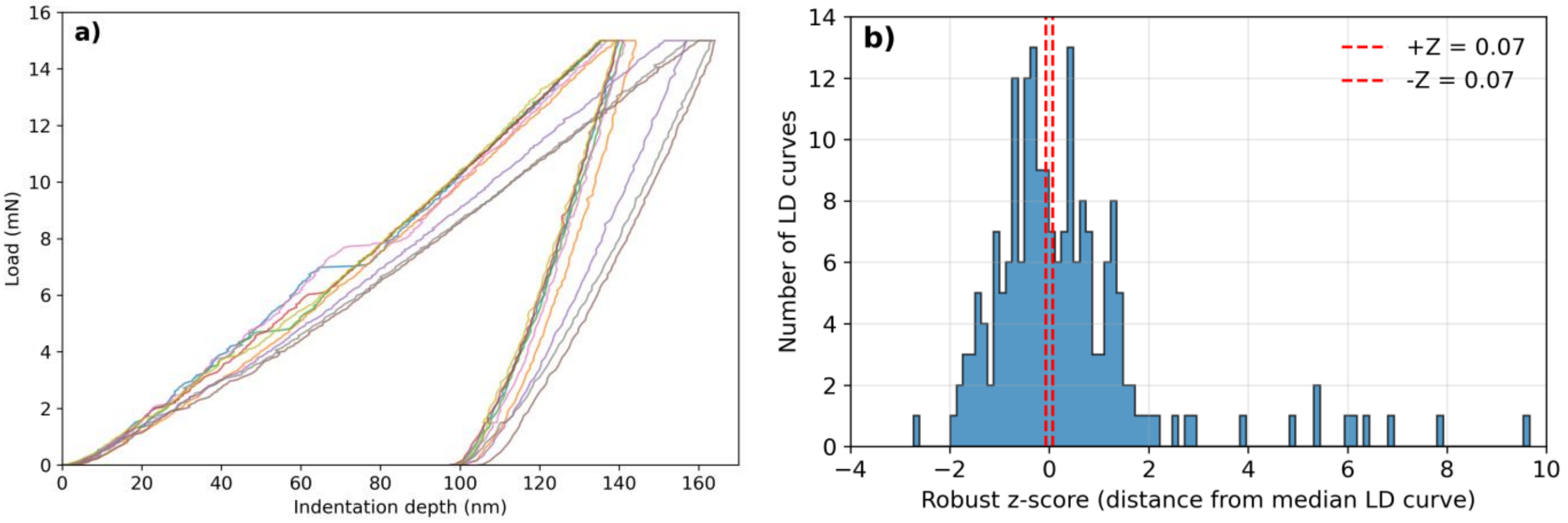

Fig 3. (a) Complete set of experimental nanoindentation load–displacement curves highlighting the selected subset of LD curves. (b) Distribution of robust z-scores quantifying the deviation of individual nanoindentation load–displacement curves from the median mechanical response. The dashed lines indicate the selected similarity threshold (Z = 0.07), defining the subset of mechanically representative curves used for comparison with molecular dynamics simulations.

Figure 4a) presents the indentation stress–strain response obtained from the representative subset of experimental load–displacement (LD) curves selected using the physics-based similarity criterion described earlier. The experimental curves exhibit a well-defined elastic regime at small indentation strains, followed by a smooth transition to plastic deformation as the contact pressure increases. The limited scatter observed within this representative set reflects the intrinsic mechanical response of the MoTaW alloy, minimizing the influence of experimental artifacts such as surface roughness, local defects, or tip–sample misalignment. This behavior confirms that the selected LD curves capture the characteristic indentation response of the material and are suitable for quantitative comparison with atomistic simulations. Figure 4b) shows the corresponding indentation stress–strain response obtained from MD simulations. A Hertzian elastic fit was applied to the low-strain regime of both the experimental and MD indentation curves to extract the reduced modulus $E_r$. The MD simulations yield a reduced modulus of $E_r^{MD} = 261.12$ GPa, while the experimental measurements give $E_r^{Exp} = 272 \pm 8$ GPa. The conversion from reduced modulus to the specimen Young's modulus $E_s$ was performed using the standard Hertzian contact-mechanics relation:

$$\frac{1}{E_r} = \frac{1-\nu_s^2}{E_s} + \frac{1-\nu_i^2}{E_i}, \qquad (2)$$

where $E_i$ and $\nu_i$ are the elastic constants of the diamond indenter, and $E_s$ and $\nu_s$ are those of the specimen. The indenter properties were taken as $E_i = 1140$ GPa and $\nu_i = 0.07$, consistent with standard nanoindentation practice. The specimen Poisson's ratio was assumed to be $\nu_s = 0.31$, based on rule-of-mixtures estimates for the equiatomic alloy. Using these inputs, the experimentally measured reduced modulus corresponds to a Young's modulus of $E_s^{Exp} = 325 \pm 13$ GPa, where the uncertainty reflects propagation of the experimental uncertainty in $E_r$. The same conversion procedure was applied consistently to the MD-derived reduced modulus for direct comparison [48]. This value lies well within the expected range bounded by the constituent refractory elements, namely Mo (~320 GPa), W (~410 GPa), and Ta (~185 GPa). The experimental uncertainty is represented by a shaded gray region, illustrating the range of elastic responses associated with measurement variability and microstructural heterogeneity. The MD-derived stress–strain curve lies well within this experimental envelope, demonstrating excellent agreement between the atomistic predictions and experimental observations in the elastic regime. This consistency validates both the contact mechanics formulation used in the simulations and the experimental data selection methodology, providing confidence in the multiscale interpretation of the nanoindentation response. The elastic modulus of MoTaW therefore falls between the softer Ta and the stiffer Mo–W end members, consistent with an approximate rule-of-mixtures–like elastic behavior. The intermediate modulus reflects the equiatomic chemical composition and the bcc crystal structure of the alloy, in which the elastic stiffness is governed primarily by bond strength and atomic packing rather than by localized chemical ordering [52]. Importantly, the relatively narrow spread between the experimental modulus and the modulus obtained from atomistic nanoindentation simulations ( 261 ± 5 GPa in terms of reduced modulus) demonstrates that the elastic response is captured consistently across length scales once the appropriate contact mechanics framework is applied. This elastic agreement provides a robust baseline for interpreting the plastic response and GSFE-controlled slip behavior discussed later, confirming that the MoTaW alloy inherits its elastic stiffness from its Mo–W-rich bonding character while remaining mechanically softer than pure tungsten.

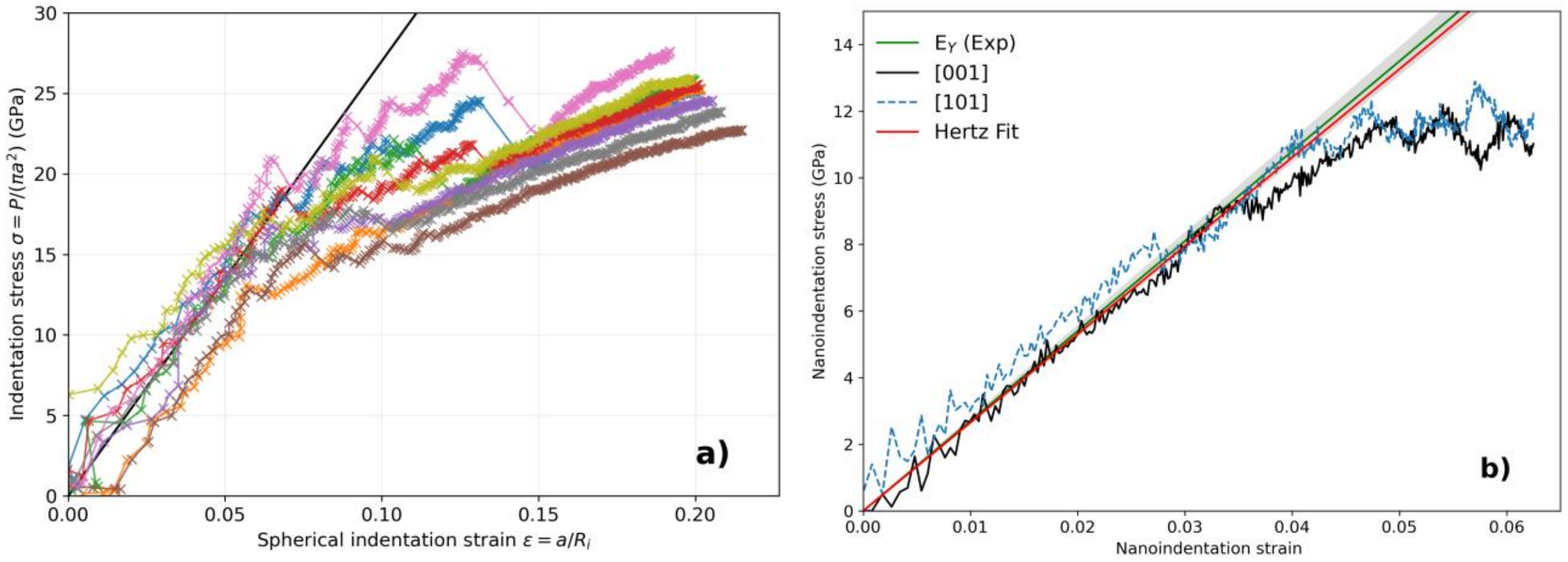

Fig 4. Indentation stress–strain response of the MoTaW alloy. (a) Stress–strain curves derived from the representative set of experimental load–displacement data. (b) Comparison between experimental and MD stress–strain responses, including Hertzian elastic fits; the shaded gray region denotes the experimental uncertainty band, highlighting the good agreement between experiment and simulation.

Beyond the purely elastic regime, the onset of incipient plasticity in the MoTaW alloy is marked by a deviation from the Hertzian elastic response and the emergence of irreversible deformation beneath the indenter. In spherical nanoindentation, this transition is commonly associated with the nucleation of the first dislocation loops once the maximum subsurface shear stress reaches a critical value. For MoTaW, the experimentally derived maximum shear stress at yielding, $\tau_{max}\approx 2.11$ GPa, occurs at a contact pressure $p_0$ consistent with the classical relationship $\tau_{max}\approx 0.31 p_0$, indicating that plasticity initiates through stress-driven dislocation nucleation rather than pre-existing defects. The relatively high stress required to trigger plasticity reflects the intrinsically high lattice resistance of the bcc refractory alloy, which is further enhanced by chemical disorder. This behavior is consistent with the GSFE trends obtained from atomistic calculations, where MoTaW exhibits elevated unstable stacking-fault and twinning-fault energies compared to Ta and Mo, implying higher energy barriers for dislocation nucleation and glide. As a consequence, the alloy sustains larger elastic strains prior to yielding, leading to a pronounced elastic-to-plastic transition. The agreement between the experimentally observed incipient plasticity and the MD simulations confirms that the initial plastic events are governed by homogeneous dislocation nucleation controlled by the underlying GSFE landscape, rather than by extrinsic microstructural features. This mechanistic link between elastic stiffness, GSFE barriers, and the onset of plasticity highlights the role of chemical complexity in stabilizing the elastic regime and delaying yielding in MoTaW.

Figure 5 illustrates the pile-up morphology formed around the nanoindentation imprint as obtained from MD simulations and experiments. Figures a) and b) show atomistic displacement-magnitude maps for indentations performed along the [001] and [011] crystallographic orientations, respectively, while c) presents a representative SEM image of an experimentally indented surface close to the [001] orientation [12, 14, 53]. For the [001] orientation (Fig. 5a), the MD results reveal a clear four-fold rosette pattern, characterized by symmetrically distributed pile-ups around the indentation site. This morphology directly reflects the activation of the dominant {110}⟨111⟩ slip systems, which are symmetrically equivalent around the [001] loading axis in bcc crystals. The pile-ups align along directions consistent with the intersection of the {110} planes with the free surface, demonstrating that plastic deformation is governed by crystallographic slip rather than isotropic flow. The pronounced lobes indicate efficient dislocation glides along these planes and limited cross-slip confinement at this indentation depth. In the case of the [011] orientation (Fig. 5b), the pile-up pattern becomes less symmetric, with an anisotropic distribution of displacement lobes. This behavior arises from the reduced symmetry of the loading direction relative to the available slip systems. While deformation is still dominated by {110} slip, the resolved shear stresses differ among equivalent planes, leading to preferential activation of a subset of slip systems and, consequently, an asymmetric rosette-like pile-up morphology [53]. This highlights the strong orientation dependence of plastic flow in the MoTaW alloy, consistent with its bcc crystal structure and GSFE-derived slip energetics. A direct comparison with the SEM image of an experimentally indented region (Fig. 5c), corresponding to a grain with surface normal indexed close to [001], shows excellent qualitative agreement with the MD predictions. The crystallographic orientation of the indented grain was determined prior to nanoindentation by EBSD mapping, allowing indentation sites to be positioned within grains of known orientation. The experimental pile-up exhibits a clear four-lobed symmetry and comparable spatial extent, closely matching the simulated rosette pattern obtained for the [001] loading direction in Fig. 5a. In the present study, we focus on representative orientations that capture the essential contrast between high-symmetry and lower-symmetry loading conditions. The [001] orientation was selected experimentally due to the availability of sufficiently large grains with well-defined surface normal alignment. An additional [011] experimental case was not included, as the objective is to demonstrate the mechanistic origin of orientation-dependent slip activation rather than to provide all crystallographic directions. The selected orientations are sufficient to establish the crystallography-controlled nature of the pile-up morphology and its correlation with the atomistic simulations. This agreement confirms that the MD simulations faithfully capture the underlying deformation mechanisms, including dislocation nucleation, glide, and surface pile-up formation. Importantly, it also validates the tabGAP MLIP's ability to reproduce crystallography-controlled plasticity under complex contact loading conditions.

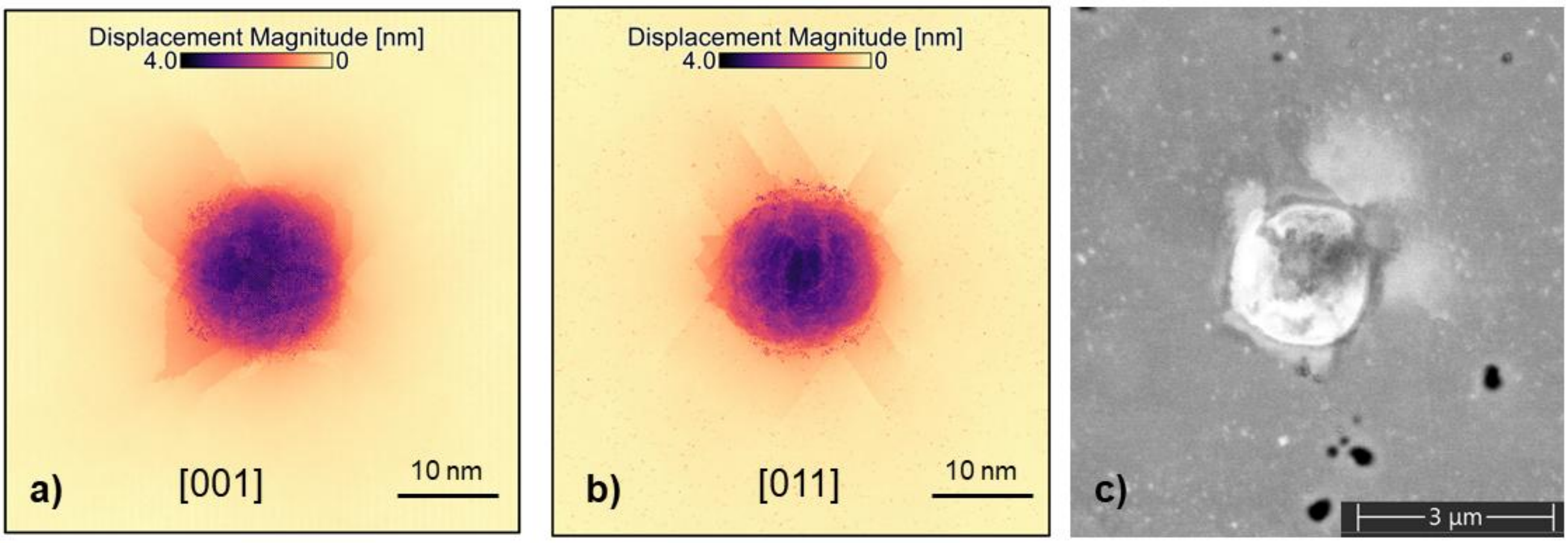

Fig 5. Pile-up morphology around nanoindentation imprints. Atomistic displacement-magnitude maps from MD simulations for indentations along (a) the [001] and (b) the [011] orientations, showing crystallography-controlled rosette patterns governed by {110} slip systems. (c) SEM image of an experimental indentation close to the [001] orientation, exhibiting a pile-up morphology in good agreement with the MD predictions.

### 3.2. Subsurface dislocation nucleation and network evolution

The subsurface plastic deformation induced by nanoindentation is governed by crystallography-dependent dislocation nucleation and glide, which are strongly influenced by the orientation of the indented surface. Figures 6(a–b) illustrate the evolution of dislocation structures beneath the indenter for the [001] and [011] orientations, respectively, revealing distinct slip activity and propagation pathways consistent with the bcc crystal symmetry. For the [001] orientation shown in Fig 6a), plastic deformation exhibits a highly symmetric response characterized by the formation of a four-fold rosette pattern. Dislocation nucleation occurs beneath the contact region and rapidly evolves into expanding half-loops that glide predominantly on the {110}⟨111⟩ slip systems, which are symmetrically equivalent around the [001] loading axis. The intersections of these {110} planes with the (001) surface project along ⟨110⟩ directions, resulting in radially distributed slip traces and a uniform propagation of dislocations away from the indent center. This symmetry facilitates homogeneous plastic flow and promotes the development of a well-organized dislocation network beneath the indenter, consistent with the relatively high unstable stacking-fault energies that delay localized shear and favor distributed slip. In contrast, the [011] orientation, depicted in Fig 6b), exhibits a markedly anisotropic dislocation response. Due to the reduced crystallographic symmetry of the loading direction, the resolved shear stresses on equivalent slip systems are no longer identical, leading to preferential activation of a subset of {110} and {112}⟨111⟩ slip systems. As a result, dislocation nucleation is spatially biased, and subsequent propagation occurs primarily along directions close to [001], producing an elongated and asymmetric plastic zone. The subsurface dislocation network in this orientation is more heterogeneous, with enhanced interaction, junction formation, and localized strain accumulation. This anisotropic behavior reflects both the orientation-dependent Schmid factors and the GSFE-derived energetic hierarchy of the available slip planes.

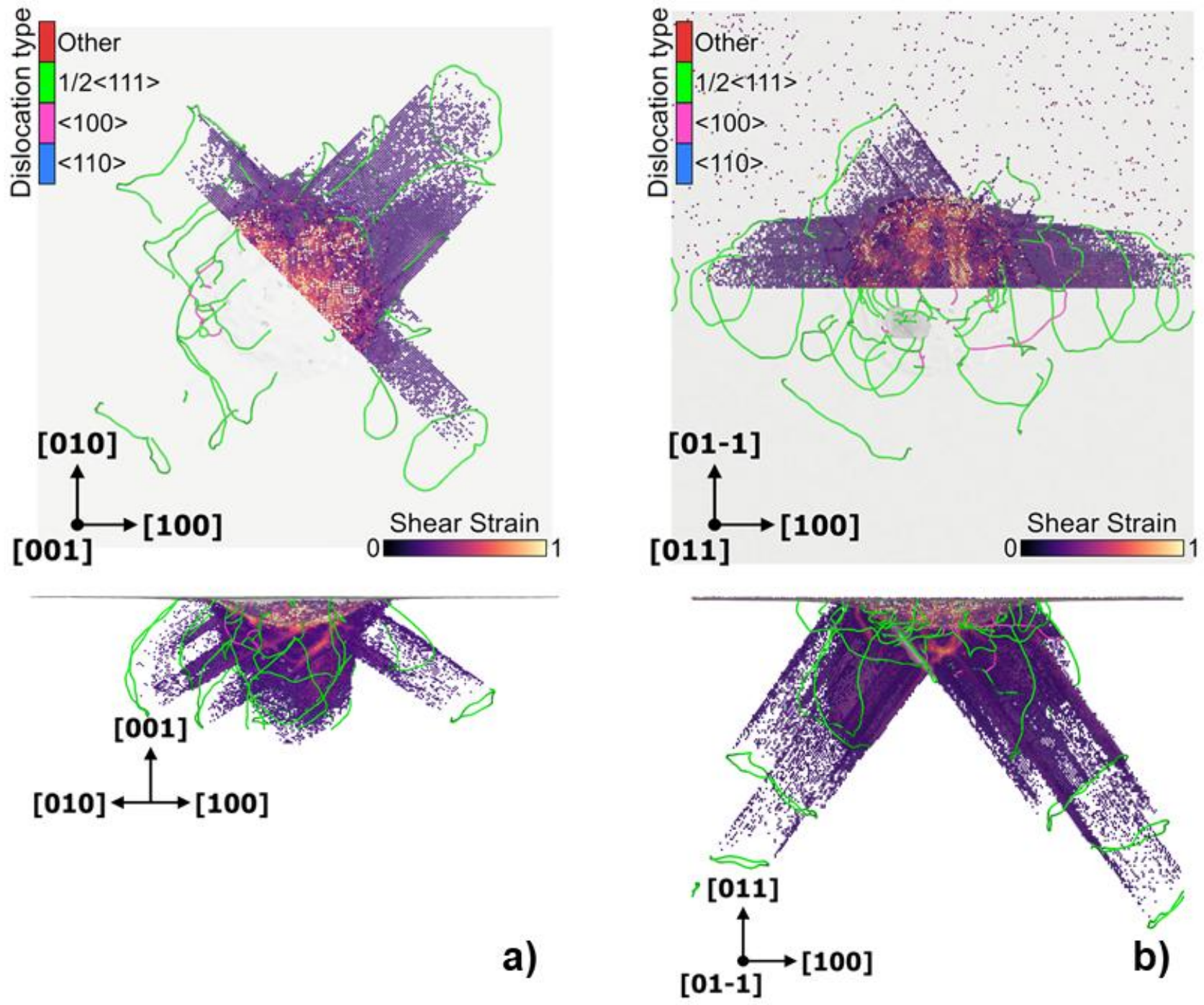

Fig 6. Subsurface dislocation structures and strain localization beneath the indenter at maximum penetration depth. Atomic strain maps overlaid with the dislocation network for the (a) [001] and (b) [011] surface orientations, illustrating orientation-dependent plastic zone development and dislocation propagation.

The evolution of the plastic zone beneath the indenter is closely linked to the crystallography-dependent slip activity and dislocation interactions discussed above. At the early stages of indentation, the strain field remains highly localized directly beneath the contact region, where the hydrostatic and shear stresses reach their maximum. As indentation proceeds and dislocation nucleation is triggered, plastic strain begins to redistribute along the active slip systems, giving rise to a growing plastic zone whose shape and extent depend strongly on the surface orientation. For the [001] orientation (Fig 6a), the symmetry of the available {110}⟨111⟩ slip systems promotes a relatively uniform redistribution of strain, resulting in a radially symmetric plastic zone with limited strain localization. The plastic region expands steadily both laterally and in depth as dislocation half-loops propagate outward and interact, forming a dense but spatially homogeneous dislocation network. In contrast, the [011] orientation (Fig 6b) exhibits pronounced strain localization due to the reduced symmetry of the resolved shear stresses on equivalent slip systems. Plastic deformation concentrates along preferential channels aligned with the dominant slip directions, leading to an elongated and anisotropic plastic zone. This localization enhances dislocation–dislocation interactions and junction formation in specific regions, promoting heterogeneous strain accumulation and a more complex subsurface network. The observed differences in plastic-zone morphology highlight the role of crystallographic orientation and GSFE anisotropy (see Fig 2) in controlling not only dislocation nucleation and glide but also the spatial distribution of plastic strain under confined contact loading.

In addition, the evolution of dislocation topology beneath the indenter plays a decisive role in the development of indentation-induced hardening in the MoTaW alloy. As plastic deformation progresses, initially isolated dislocation half-loops expand and increasingly interact, giving rise to a dense three-dimensional network characterized by frequent dislocation intersections and <100> type dislocation junction formation, between 1/2[1-1-1] and 1/2[-11-1] dislocations.

For the [001] orientation, the simultaneous activation of multiple equivalent {110}⟨111⟩ slip systems leads to extensive dislocation interactions and the formation of forest dislocations that impede further glide. The resulting network is dominated by screw and mixed-character segments capable of cross-slip, as shown in Fig 6a), which enhances network connectivity while progressively increasing the resistance to dislocation motion. This mechanism is characteristic of forest hardening, where mobile dislocations are obstructed by pre-existing dislocation lines, leading to a gradual increase in flow stress under continued indentation. For the [011] orientation (Fig 6b), the reduced symmetry of the resolved shear stresses restricts slip activity to fewer preferential systems, promoting the accumulation of dislocations along specific directions. This anisotropic slip activity enhances the likelihood of junction formation and localized dislocation tangles, which act as strong obstacles to subsequent dislocation motion. These junctions effectively immobilize segments of the dislocation network, intensifying strain localization and contributing to a more pronounced hardening response. The orientation-dependent differences in dislocation topology thus directly control the balance between forest hardening and junction-controlled hardening mechanisms. These atomistic observations provide a mechanistic explanation for the orientation sensitivity of the nanoindentation response and demonstrate how GSFE-governed slip activity dictates the evolution of dislocation networks and the associated hardening behavior under confined contact loading.

Local structural entropy provides a physically motivated scalar descriptor to quantify deviations from the ideal crystalline environment at the atomic scale [54]. Thus, the local structural disorder was quantified using the local entropy (pair-entropy) descriptor as implemented in OVITO. The radial integration cutoff was set to 5.0 Å, extending beyond the first-neighbor shell and approaching the first minimum of the radial distribution function of the reference BCC structure, thereby ensuring adequate characterization of the local atomic environment. A Gaussian smoothing parameter of 0.20 Å was employed to reduce thermal noise while preserving defect-induced structural variations. Local-density normalization and neighbor averaging were enabled to improve robustness against hydrostatic strain beneath the indenter and to obtain a spatially stable representation of structural disorder. These parameters were selected to ensure reliable and reproducible identification of lattice distortions and defect structures during indentation. We compute the local entropy by using the two-body excess entropy formulation, in which the local radial distribution function (g(r)) around each atom is evaluated within a finite cutoff radius and compared to that of a perfectly ordered reference which is the sample before nanoindentation test. The polyhedral template matching (PTM) method implemented in OVITO identifies the local crystal structure of each atom by comparing its neighbor topology and geometry to ideal reference templates (e.g., bcc, fcc, hcp, etc.). In practice, PTM constructs the local coordination polyhedron from a chosen neighbor shell, performs an optimal mapping between the measured neighbor arrangement and each template, and assigns the structure that minimizes the geometric misfit (often expressed through a residual/deformation metric). Because the matching relies on both connectivity and local geometry, PTM is less sensitive to thermal noise than purely distance-based metrics and can robustly classify atoms in the presence of elastic strain, making it well suited for indentation simulations where strong stress gradients develop beneath the indenter. In Fig 7, we present the resulting entropy measurements for orientations [001] in a) and [011] in b), reflecting the degree of local structural disorder: atoms in an ideal bcc environment exhibit strongly negative entropy values, while atoms experiencing lattice distortion, coordination changes, or bond rearrangements show progressively less negative values; these atoms are typically located in the region of close contact with the indenter tip and where the maximum stress is located at 0.5 nm underneath the indenter tip region, and associated to dislocation cores. Dislocation lines appear as continuous filaments of elevated local entropy embedded within the crystalline matrix, while regions of high dislocation density and junction formation are highlighted by broader zones of increased entropy. Compared to purely topological defect identifiers, local entropy captures both elastic distortions and plastic rearrangements, enabling a unified visualization of the evolving dislocation network and plastic zone development.

Applying PTM to the nanoindentation simulations reveals that plastic deformation is accompanied by localized changes of atomic-level symmetry in the highly sheared contact zone. Although the bulk matrix remains bcc, atoms within the evolving plastic zone—particularly near dislocation cores, junctions, and the highly strained region beneath the tip, are frequently classified as fcc- or hcp-like due to the transient and/or stabilized local stacking rearrangements associated with intense shear and non-affine deformation. At the maximum indentation depth, for both the [001] and [011] orientations, the population of atoms not identified as bcc is dominated by close-packed motifs: 31% are classified as fcc and 69% as hcp, indicating that the non-bcc fraction is overwhelmingly associated with locally close-packed environments generated by deformation rather than by long-range phase transformation. Importantly, tracking the chemical identity of these non-bcc atoms confirms that the Mo, Ta, and W populations remain equiatomic, demonstrating that the observed structural changes arise from deformation-driven local rearrangements rather than compositional segregation or element-specific phase preference.

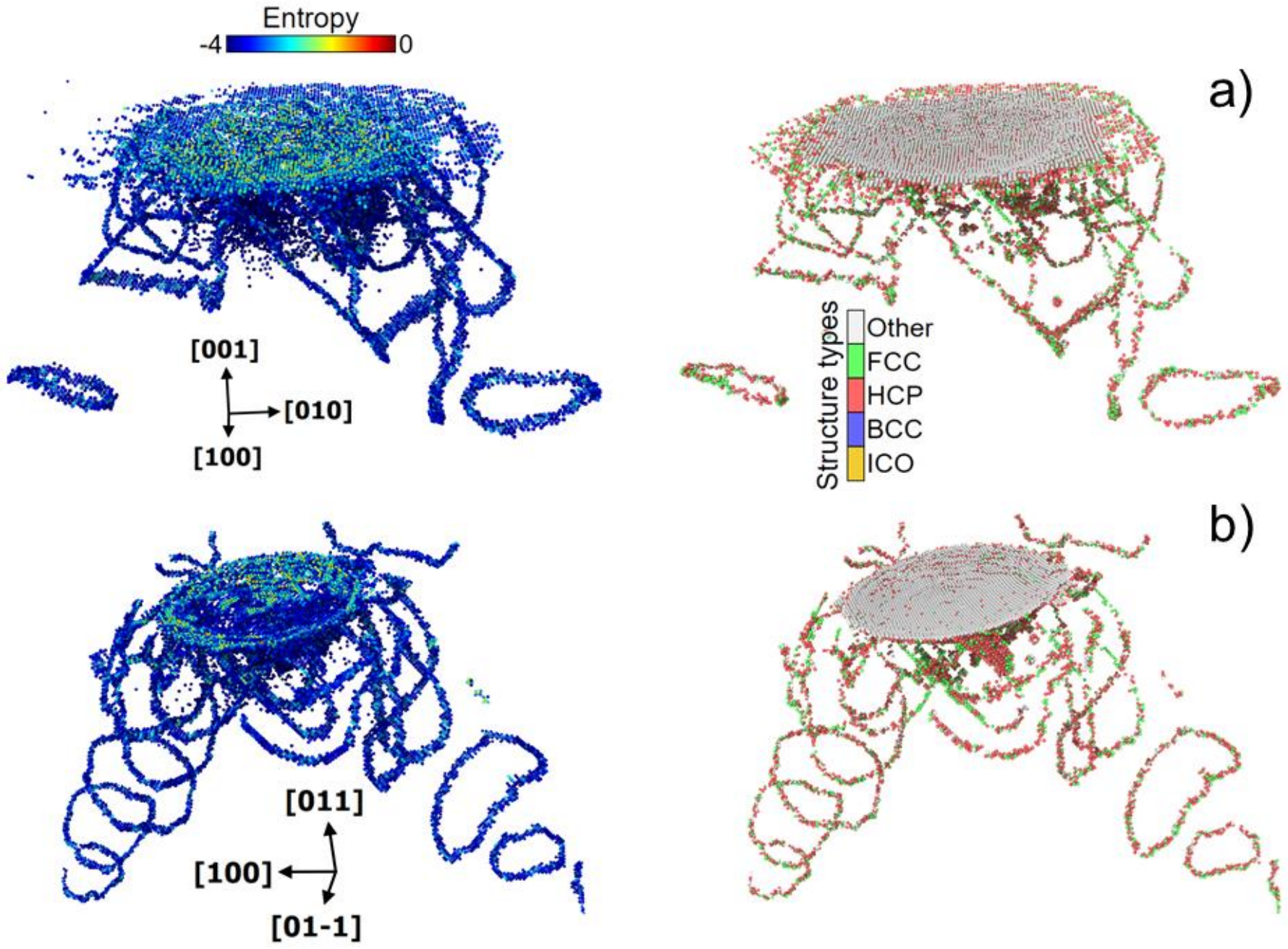

Figure 7. Subsurface deformation beneath the indenter at maximum penetration depth. Left panels show the spatial distribution of local atomic entropy, highlighting regions of intense non-affine deformation and dislocation activity, while right panels display the corresponding local crystal structure identified by the polyhedral template matching (PTM) method. Results are shown for (a) the [001] and (b) the [011] loading orientations, illustrating orientation-dependent dislocation network morphology and associated structural transformations beneath the contact.

## 4. Conclusion

In this work, experimentally guided molecular dynamics simulations with nanoindentation experiments are combined to elucidate the deformation mechanisms governing the mechanical response of an equiatomic MoTaW refractory complex concentrated alloys. Generalized stacking fault energy calculations revealed that chemical disorder in MoTaW leads to non-linear fault energetics, with unstable stacking-fault and unstable twinning-fault energies that exceed simple rule-of-mixtures predictions. These elevated energy barriers indicate a suppression of localized shear and deformation twinning, favoring distributed dislocation-mediated plasticity at the onset of deformation. Large-scale nanoindentation simulations using a tabulated low-dimensional Gaussian Approximation Potential (tabGAP) captured orientation-dependent plasticity with high fidelity. The simulated load–displacement responses and indentation stress–strain curves show excellent agreement with experimentally derived data, particularly in the elastic regime, where the reduced Young's modulus obtained from MD (261.1 GPa) lies within the experimental uncertainty band (273 ± 8 GPa). This agreement validates both the contact-mechanics framework employed and the accuracy of the tabGAP potential for describing elastic and incipient plastic behavior in refractory multicomponent alloys. Atomistic analyses of surface pile-up morphologies and subsurface dislocation structures revealed that plastic deformation is strongly controlled by crystallographic orientation. For the [001] orientation, symmetric activation of {110}⟨111⟩ slip systems leads to four-fold rosette pile-up patterns and a relatively homogeneous plastic zone. In contrast, the [011] orientation exhibits anisotropic slip activity, enhanced strain localization, and a more heterogeneous dislocation network. The evolution of dislocation topology, including forest hardening and junction formation, was shown to play a central role in governing indentation-induced hardening, with distinct mechanisms emerging depending on the symmetry of the activated slip systems.

The local entropy and PTM analyses provide complementary insights into the subsurface plasticity mechanisms activated during nanoindentation of bcc MoTaW. Regions of elevated local entropy are concentrated beneath the indenter and along curved subsurface features, directly correlating with the nucleation and propagation of dislocation networks. These high-entropy zones mark areas of strong non-affine atomic rearrangements, identifying the evolving plastic zone with greater sensitivity than conventional strain measures. The PTM results reveal that atoms departing from the bcc reference structure predominantly transform into FCC- and HCP-like local environments, which decorate the dislocation cores and junctions. For both the [001] and [011] loading orientations at maximum indentation depth, approximately 31% of the non-bcc atoms are classified as FCC and 69% as HCP, indicating that HCP-like configurations dominate the defected regions. This structural partitioning is characteristic of dissociated dislocations and complex junctions in bcc systems under high contact stresses. Importantly, despite these local structural transformations, the chemical composition within the plastically deformed zone remains equiatomic in Mo, Ta, and W, confirming that the observed changes are mechanically driven rather than compositionally induced. Together, these results highlight how orientation-dependent dislocation topology governs strain localization and defect evolution during indentation in refractory multi-principal-element alloys. Overall, this study establishes a direct mechanistic link between GSFE-derived fault energetics, orientation-dependent dislocation activity, and experimentally observed nanoindentation behavior. The results demonstrate the capability of tabGAP-based simulations to bridge atomistic mechanisms and experimental observations, providing a robust multiscale framework for understanding and predicting plasticity in refractory complex concentrated alloys under contact loading.

**Acknowledgments**

Research was funded through European Union Horizon 2020 research and innovation program under Grant Agreement No. 857470 and from the European Regional Development Fund under the program of the Foundation for Polish Science International Research Agenda PLUS, Grant No. MAB PLUS/2018/8, and the initiative of the Ministry of Science and Higher Education "Support for the activities of Centers of Excellence established in Poland under the Horizon 2020 program" under Agreement No. MEiN/2023/DIR/3795. The Ministry of Science, Technological Development, and Innovation of the Republic of Serbia, grant No. 451-03-136/2025-03/200023. We gratefully acknowledge Polish high-performance computing infrastructure PLGrid (HPC Center: ACK Cyfronet AGH) for providing computer facilities and support within computational Grant No. PLG/2024/017084. The experimental part of the research was also partially funded by the National Science Centre (NCN), grant number MINIATURA-7 2023/07/X/ST11/00862.

## References

[1] Senkov ON, Gorsse S, Miracle DB. High temperature strength of refractory complex concentrated alloys. Acta materialia. 175, 394 (2019).
[2] Wang, W., Kumar, P., Cook, D.H. et al. Ductility mechanisms in complex concentrated refractory alloys from atomistic fracture simulations. npj Comput Mater 11, 330 (2025).
[3] Fan, X., Qu, R. & Zhang, Z. Remarkably high fracture toughness of hfnbtatizr refractory high-entropy alloy. J. Mater. Sci. Technol. 123, 70 (2022).
[4] Tomasz Stasiak, Pavel Souček, Vilma Buršíková, Nikola Koutná, Zsolt Czigány, Katalin Balázsi, Petr Vašina. Synthesis and characterization of the ceramic refractory metal high entropy nitride thin films from Cr-Hf-Mo-Ta-W system. Surface and Coatings Technology 449, 128987 (2022).
[5] Tomasz Stasiak, Stanislava Debnarova, Shuyao Lin, Nikola Koutna, Zsolt Czigany, Katalin Balazsi, Vilma Buršíková, Petr Vašina, Pavel Souček. Synthesis and characterization of ceramic high entropy carbide thin films from the Cr-Hf-Mo-Ta-W refractory metal system. Surface and Coatings Technology 485, 130839 (2024).
[6] Miracle DB, Senkov ON, Frey C, Rao S, Pollock TM. Strength vs temperature for refractory complex concentrated alloys (RCCAs): A critical comparison with refractory BCC elements and dilute alloys. Acta Materialia 266, 119692 (2024).
[7] Belcher CH, Kamp D, To S, Lu Y, Chassaing D, Boll T, MacDonald BE, Lee EM, Apelian D, Lavernia EJ. The origin and control of interstitial impurities in refractory complex concentrated alloys. Journal of Alloys and Compounds 1010, 177520 (2025).
[8] Butler TM, Chaput KJ, Dietrich JR, Senkov ON. High temperature oxidation behaviors of equimolar NbTiZrV and NbTiZrCr refractory complex concentrated alloys (RCCAs). Journal of Alloys and Compounds. 729, 1004 (2017).
[9] Zhang Y, Xu Z, Zhang Z, Yao W, Hui X, Liang X. Microstructure and mechanical properties of Mo-Ta-W refractory multi-principal element alloy thin films for hard protective coatings. Surface and Coatings Technology 431, 128005 (2022).
[10] Deringer VL, Caro MA, Csányi G. Machine learning interatomic potentials as emerging tools for materials science. Advanced Materials 31, 902765 (2019).
[11] F. J. Domínguez-Gutiérrez, P. Grigorev, A. Naghdi, J. Byggmästar, G. Y. Wei et al. Nanoindentation of tungsten: From interatomic potentials to dislocation plasticity mechanisms. Phys. Rev. Materials 7, 043603 (2023).
[12] F. J. Domínguez-Gutiérrez, A. Olejarz, M Landeiro Dos Reis, E Wyszkowska, D Kalita et al. Atomistic-level analysis of nanoindentation-induced plasticity in arc-melted NiFeCrCo alloys: The role of stacking faults. J. Appl. Phys. 135, 185101 (2024)
[13] F. J. Dominguez-Gutierrez, S. Papanikolaou, S. Bonfanti, M Alava. Plastic deformation mechanisms in BCC single crystals and equiatomic alloys: Insights from nanoindentation. Computer Methods in Materials Science 24), 37-49 (2024).

[14] Varillas, J., Očenášek, J., Torner, J., & Alcalá, J.. Unraveling deformation mechanisms around FCC and BCC nanocontacts through slip trace and pileup topography analyses. Acta Materialia 125, 431 (2017).
[15] F. J. Dominguez-Gutierrez, M. Frelek-Kozak, G. Markovic, M. A. Stróżyk, A. Daramola et al. High-temperature deformation behavior of Co-free nonequiatomic CrMnFeNi alloy. Phys. Rev. Materials 9, 123607 (2025).
[16] A. Daramola, G. Bonny, G. Adjanor, C. Domain, G. Monnet, and A. Fraczkiewicz, Comput. Mater. Sci. 203, 111165 (2022).
[17] Behler J. Perspective: Machine learning potentials for atomistic simulations. The Journal of chemical physics 145, 170901 (2016).
[18] Anstine DM, Isayev O. Machine learning interatomic potentials and long-range physics. The Journal of Physical Chemistry A. 127, 2417 (2023).
[19] Wu L, Li T. A machine learning interatomic potential for high entropy alloys. Journal of the Mechanics and Physics of Solids 187, 105639 (2024).
[20] Byggmästar J, Nordlund K, Djurabekova F. Modeling refractory high-entropy alloys with efficient machine-learned interatomic potentials: Defects and segregation. Phys. Rev. B 104, 104101 (2021).
[21] Fellman A, Byggmästar J, Granberg F, Djurabekova F, Nordlund K. Radiation damage and phase stability of Alx CrCuFeNiy alloys using a machine-learned interatomic potential. Phys. Rev. Materials 10, 023801 (2026)
[22] Byggmästar J, Nordlund K, Djurabekova F. Simple machine-learned interatomic potentials for complex alloys. Physical Review Materials 6, 083801 (2022).
[23] Xueqin Pang,Xianglin Mu, Zilei Bai, Jiayun Deng, Xiaoning Wen et al. Atomic-scale study of ultrasonic elliptical vibration-assisted cutting of InP crystals via molecular dynamics. Journal of Materials Research and Technology 41, 2275 (2026).
[24] K. Mulewska, D. Kalita, M. Wilczopolska, W. Chromiński, P.A. Ferreirós, Ł. Kurpaska. Microstructural evolution and mechanical response of ion-irradiated Fe-9Cr alloys: Insights from nanoindentation. Journal of Nuclear Materials 616, 156109 (2025).
[25] J. Byggmästar, K. Nordlund, and F. Djurabekova. Gaussian approximation potentials for body-centered-cubic transition metals. Phys. Rev. Materials 4, 093802 (2020).
[26] Albert P Bartók, Mike C Payne, Risi Kondor, Gábor Csányi. Gaussian Approximation Potentials: The Accuracy ofQuantum Mechanics, without the Electrons. Phys. Rev. Lett. 104, 136403 (2010).
[27] Ryan Jacobs, Dane Morgan, Siamak Attarian, Jun Meng, Chen Shen, Zhenghao Wu, Clare Yijia Xie, Julia H Yang, Nongnuch Artrith, Ben Blaiszik et al. Current Opinion in Solid State and Materials Science 35, 101214 (2025)
[28] Sergei Starikov, Petr Grigorev, Pär A.T. Olsson. Angular-dependent interatomic potential for large-scale atomistic simulation of W-Mo-Nb ternary alloys. Computational Materials Science 233, 112734 (2024).
[29] Haoyu Hu, Chao Zhang, Rui Yue, Biao Hu, Shuai Chen. Unraveling ductility enhancement mechanisms in W-Ta alloys using machine-learning potential. International Journal of Mechanical Sciences 286, 109911 (2025).
[30] T. Swinburne and C. Marinica. Unsupervised Calculation of Free Energy Barriers in Large Crystalline Systems. Phys. Rev. Lett. 120, 135503 (2018).
[31] Xiaowang Wang, Shuozhi Xu, Wu-Rong Jian, Xiang-Guo Li, Yanqing Su, Irene J. Beyerlein. Generalized stacking fault energies and Peierls stresses in refractory body-centered cubic metals from machine learning-based interatomic potentials. Computational Materials Science 192, 110364 (2021).
[32] FJ Domínguez-Gutiérrez, S Papanikolaou, A Esfandiarpour, P Sobkowicz, M Alava. Nanoindentation of single crystalline Mo: Atomistic defect nucleation and thermomechanical stability. Materials Science and Engineering: A 826, 141912 (2021).
[33] Guénolé, J., Nöhring, W. G., Vaid, A., Houllé, F., Xie, Z., Prakash, A., and Bitzek, E.. Assessment and optimization of the fast inertial relaxation engine (FIRE) for energy minimization in atomistic simulations and its implementation in LAMMPS. Computational Materials Science, 175, 109584 (2020).
[34] A. P. Thompson, H. M. Aktulga, R. Berger, D. S. Bolintineanu, W. M. Brown, P. S. Crozier, P. J. in 't Veld, A. Kohlmeyer, S. G. Moore, T. D. Nguyen, R. Shan, M. J. Stevens, J. Tranchida, C. Trott, and S. J. Plimpton, LAMMPS - a flexible simulation tool for particle-based materials modeling at the atomic, meso, and continuum scales. Comput. Phys. Commun. 271, 108171 (2022).
[35] Stukowski, A.. Visualization and analysis of atomistic simulation data with OVITO-the Open Visualization Tool. Modelling and Simulation in Materials Science and Engineering 18(1), 015012 (2010).
[36] Stukowski, A., Bulatov, V. V., & Arsenlis, A.. Automated identification and indexing of dislocations in crystal interfaces. Modelling and Simulation in Materials Science and Engineering, 20(8), 085007 (2012).
[37] Jing Qian, C.Y. Wu, J.L. Fan, H.R. Gong. Effect of alloying elements on stacking fault energy and ductility of tungsten. Journal of Alloys and Compounds 737, 372 (2018).
[38] X.W. Zhou, R.A. Johnson, H.N.G. Wadley, Misfit-energy-increasing dislocations in vapor-deposited CoFe/NiFe multilayers, Phys. Rev. B 69, 144113 (2024).

[39] Jorge Alcalá, Roger Dalmau, Oliver Franke, Monika Biener, Juergen Biener, and Andrea Hodge. Planar Defect Nucleation and Annihilation Mechanisms in Nanocontact Plasticity of Metal Surfaces. Phys. Rev. Lett. 109, 075502 (2012).
[40] Mamun AA, Xu S, Li XG, et al. Comparing interatomic potentials in calculating basic structural parameters and Peierls stress in tungsten-based random binary alloy. Phys Scr 98, 105923 (2023).
[41] Wei N, Jia T, Zhang X, et al. First-principles study of the phase stability and the mechanical properties of W-Ta and W-Re alloy. AIP Adv 4, 057103 (2014).
[42] Scikit-learn: Machine Learning in Python, Pedregosa et al., JMLR 12, 2825, (2011).
[43] Edoardo Rossi, Daniele Duranti, Saqib Rashid, Michal Zitek, Rostislav Daniel, Marco Sebastiani. A high-throughput framework for pile-up correction in high-speed nanoindentation maps. Materials & Design 251, 113708 (2025).
[44] A. Bolshakov, G.M. Pharr. Influences of pileup on the measurement of mechanical properties by load and depth sensing indentation techniques. J. Mater. Res., 13, 1049 (2011).
[45] G.M. Pharr. Measurement of mechanical properties by ultra-low load indentation. Materials Science and Engineering: A 253, 151 (1998)
[46] Brian R. Law. Indentation of Ceramics with Spheres: A Century after Hertz. J. Am. Ceram. Soc. 81, 1977 (1998)
[47] A.C. Fischer-Cripps. Critical review of analysis and interpretation of nanoindentation test data. Surface and Coatings Technology 200, 4153 (2006).
[48] Pathak, S., & Kalidindi, S. R. Spherical nanoindentation stress–strain curves. Materials Science and Engineering: R: Reports, 91, 1–36 (2015).
[49] Pathak, S., Riesterer, J. L., Kalidindi, S. R., & Michler, J.. Understanding pop-ins in spherical nanoindentation. Applied Physics Letters 105, 161913 (2014)
[50] W.C. Oliver, G.M. Pharr. Measurement of hardness and elastic modulus by instrumented indentation: advances in understanding and refinements to methodology. J. Mater. Res., 19, 3-20 (2011).
[51] M. Zhao, N. Ogasawara, N. Chiba, X. Chen. A new approach to measure the elastic–plastic properties of bulk materials using spherical indentation. Acta Materialia 54, 23 (2006)
[52] Liu, G., Zhang, G. J., Jiang, F., Ding, X. D., Sun, Y. J., Sun, J., & Ma, E.. Nanostructured high-strength molybdenum alloys with unprecedented tensile ductility. Nature Materials, 12, 344 (2013).
[53] Hongbing Yu, Suchandrima Das, Haiyang Yu, Phani Karamched, Edmund Tarleton, Felix Hofmann. Orientation dependence of the nano-indentation behaviour of pure Tungsten. Scripta Materialia 189, 135-139 (2020).
[54] P.M. Piaggi and M. Parrinello. Entropy based fingerprint for local crystalline order, J. Chem. Phys. 147, 114112 (2017).